\documentstyle[epsfig]{article}

\def\m{\mu}

\def\~{\widetilde}

\def\bY3{\bar Y_{,3}}
\def\Y3{Y_{,3}}

\def\Y{{\bar Y}}

\def\`{\dot}
\def\be{\begin{equation}}
\def\ee{\end{equation}}
\def\bea{\begin{eqnarray}}
\def\eea{\end{eqnarray}}

\def\fn{\footnote}

\def\mn{{\mu\nu}}

\begin{document}

\title{Renormalization by Gravity and the Kerr spinning particle}

\author{Alexander Burinskii\\
Gravity Research Group, NSI Russian
Academy of Sciences\\
B. Tulskaya 52, 115191 Moscow, Russia}
\maketitle

\begin{abstract}
On the basis of the Kerr spinning particle, we show that the mass
renormalization is perfectly performed by gravity for an arbitrary
distribution  of source matter. A smooth regularization of the
Kerr-Newman solution is considered, leading to a source in the form
of a rotating bag filled by a false vacuum. It is shown that gravity
controls the phase transition to an AdS or dS false vacuum state inside
the bag, providing the mass balance.
\end{abstract}

\section{Introduction}
Many of complicated quantum procedures admit an
interpretation in terms of some classical analogues.
The mass renormalization in QED represents a peculiar case.
It is universally recognized due to an incredible exactness of
its predictions, and its origin
lies in the classical theory of a pointlike electron;
nevertheless, there are serious problems with physical
interpretation and mathematical correctness of this procedure.

In this paper, we consider a semiclassical model of a spinning particle
based on the Kerr-Newman solution of the Einstein-Maxwell theory.
This solution has double gyromagnetic ratio,
as that of the Dirac electron and may be considered as a model of electron
in general relativity \cite{Car,DKS,Bur,Lop}.

In this paper we would like to show that the mass renormalization and
regularization of the singularities in the Kerr-Newman source are perfectly
realized by gravitational field in a very natural manner.
It allows one to conjecture that the methodological problems in QED may be
related to the ignorance of gravity.
QED ignores gravitational field arguing that
its local action is negligible. It is true, but only
partially. The Kerr solution gives a contr-example to this assertion,
showing that the local action of the
gravitational field may extend on the Compton distances due to the
stringy structure of the source.
However, the main effect of gravity is apparently related to a non-local
action. We would like to show here that in the semiclassical model of
the Kerr spinning particle, gravity provides the mass renormalization.

\section{Renormalization by gravity}

The mass of an isolated source
is determined only by an asymptotic gravitational field, and, therefore, it
depends only on the mass parameter $m$ which survives in the
asymptotic expansion of the metric.
On the other hand, the total mass can be calculated as a volume integral,
which takes into account densities of the electromagnetic energy
$\rho_{em},$ material (mechanical mass) sources $\rho_{m}$
and energy of the gravitational field $\rho_{g}$.
The last term is not taken into account in QED, but it
provides perfect renormalization.
For a spherically symmetric system, the expression may be reduced to an
integral over radial distance $r ,$

\be m = 4\pi \int_0^\infty \rho_{em} dr
+ 4\pi\int_0^\infty \rho_{m} dr + 4\pi\int_0^\infty \rho_{g} dr
\label{mtot} .
\ee
It looks like the expressions in a flat spacetime.
However, in the Kerr-Schild background it is a consequence of the exact Tolman
relations taking into account energy of matter, energy of gravitational
field (including the contribution from pressure) and rotation  \cite{BEHM}.
In the well known classical model of an electron  as a charged
sphere with electromagnetic radius $r_e =\frac {e^2}
{2m},$  integration in (\ref{mtot}) is performed in the
diapason $[r_0, \infty],$ where $r_0=r_e.$ The total mass is
determined by electromagnetic contribution only, and  contribution
 from gravity turns out to be null. However, if
$r_0<r_e,$ the electromagnetic  contribution exceeds the total mass and
this extension is to be compensated by the negative gravitational contribution.
Indeed, the results will not depend on the cut parameter $r_0$ and,
moreover, on radial distribution of matter at all.
 Some of the terms may  be divergent, but the total
result will not be changed, since divergences will always be
compensated
by a contribution from a gravitational term.

It shows that, due to the strong non-local action,
gravity may be essential for elementary particles,
on the distances which are very far from the usually considered
Planck scale.

\section{Structure of the Kerr geometry}

The Kerr-Newman solution breaks the prevailing point of view that
the local
action of gravitational field of a particle extends to its
Schwarzschild radius. The Schwarzschild singular point turns in
the Kerr rotating  geometry into a singular ring which extends
on the Compton sizes, since its radius $a=J/m$,
for $J\sim \hbar$, is the Compton one, which exceeds the
Schwarzschild one for an electron at $\sim 10^{22}$.
Angular momentum $J=\hbar /2$ for parameters of
electron is so high that the black hole horizons disappear, and the
source of the Kerr spinning particle represents a naked singular
ring which may have some stringy excitations, generating the spin and
mass of the extended particle-like object - ``microgeon'' \cite{Bur}.
Therefore, the Kerr source represents a closed singular string of
 the Compton size, and cannot be localized in the region which is
 smaller then the Compton size. \fn{In this respect the Kerr source is
 similar to the Dirac wave function.} It was shown, that this source is
 indeed a string \cite{Bur,BurAxi,BurOri} resembling a heterotic string of
 superstring theory. \fn{Adding a
stringy tension $T$ to the Kerr source, $E\equiv m =Ta$, and
combining this relation with $J=ma$, one obtains the Regge dependence
$J=\frac 1T m^2$.}
Note that this singularity is
 a branch line of the Kerr space which turns out to be two-sheeted,
 and the disk spanned on this ring plays the role of gates to
 anti-world (``negative'' sheet), where the signs of charges and
 masses, and the directions of the fields are changed.
 \fn{It was discussed many times, see for
 example \cite{BurOri,BurNst}.} So, the Kerr string is an ``Alice'' one,
 and all the fields have to fill these `gates to anti-world'
 which have the giant Compton sizes ($\sim 10^{-11}$ cm).
 Note that in QED it is the region of virtual
 photons.

One more remarkable structure of the Kerr geometry is PNC (principal
null congruence). It is a vortex of the lightlike rays (twistors)
which fall on the `negative sheet' on the Kerr disk,
penetrate it and turn into outgoing `out'-fields on the
`positive sheet' of space (see fig.1). PNC is a very important
object since the tangent to congruence vector $k^\m$ determines
the Kerr-Schild ansatz for metric
\be g^\mn =\eta^\mn + 2H k^\m k^n \label{KS} \ee (where $\eta^\mn$
is the auxiliary Minkowski metric) and
the form of vector potential
\be A_\m = {\cal A} (x) k_\m \label{Aem} \ee
for electrically charged solution, i.e. it determines polarization of
the gravitational and
electromagnetic fields around the Kerr source and the directions of
radiation for the nonstationary excited solutions \cite{BurAxi,BurNst}.

\begin{figure}[ht]
\centerline{\epsfig{figure=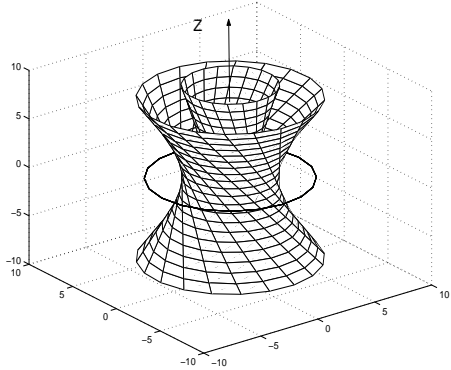,height=5cm,width=6cm}}
\caption{The Kerr singular ring and 3-D section of the Kerr principal null
congruence (PNC). Singular ring is a branch line of space, and PNC
propagates from ``negative'' sheet of the Kerr space to ``positive '' one,
covering the space-time twice. } \end{figure}

The Kerr congruence is  determined by
the Kerr theorem \cite{BurNst,KraSte,Multiks} in terms of twistors.
 The Kerr singular ring is a focal line of the Kerr PNC.

The Kerr-Schild form of metric allows one to
consider a broad class of regularized solutions which
remove the Kerr singular ring, covering it by a matter
source. There is a long-term story of the attempts to find
some interior regular solution for the
Kerr or Kerr-Newman solutions \cite{Car,Lop,BurBag,BEHM}.
\fn{Extra references may be found in \cite{BurBag,BEHM}.}
Usually, the regularized solutions have to retain
the Kerr-Schild form of metric
and the form of Kerr principal null congruence $k_\m(x),$
as well as its property
to be geodesic and shear-free.
The space part $\vec n$ of the Kerr congruence $k_\m=(1,\vec n)$
has the form of a spinning hedgehog.  Indeed, by setting the parameter
of rotation $a$ equal to zero, the Kerr singular ring shrinks to
a singular point, and $\vec n$ takes the usual hedgehog form
which is used as an ansatz for the solitonic
models of elementary particles and quarks.
It suggests that the Kerr spinning particle may have relation not only to
electron, but also to the other
elementary particles. Indeed, the Kerr-Schild class
of metric has a remarkable property, allowing us to consider a broad class of
the charged and uncharged, the spinning and spinless solutions from an unified
point of view.
\section{Regularization of the Kerr singularity}

Our treatment will be based on the approach given in
\cite{BurBag,BEHM}, where the {\it smooth}
regularized sources were obtained for the rotating and
non-rotating solutions of the Kerr-Schild class.
These smooth and regular solutions have the scalar function $H$
of the general form
\be
H=f(r)/(r^2 + a^2 \cos^2 \theta) \label{hf}.
\ee
For the Kerr-Newman solution function $f(r)$ has
the form
\be
f(r)\equiv f_{KN}= mr -e^2/2 \label{hKN}.
\ee

Regularized solutions have tree regions:

i) the Kerr-Newman exterior, $r>r_0 $, where $f(r)=f_{KN} ;$

ii) interior $r<r_0-\delta $, where $f(r) =f_{int}$ and
function $f_{int}=\alpha r^n ,$ and $n\ge 4$ to suppress
the singularity at $r=0,$ and provide the smoothness of the metric
up to the second derivatives;

iii) a narrow intermediate region $r\in [r_0-\delta, r_0]$ which
allows one to get a smooth solution interpolating between regions
i) and ii).

It is advisable to consider first the non-rotating cases, since the
rotation can later be taken into account by an easy trick. In this
case, taking $n=4$ and the parameter $\alpha=8\pi \Lambda/6 ,$ one
obtains for the source (interior) a space-time of  constant
curvature $R=-24 \alpha$ which is generated by a source with
energy density

\be \rho = \frac 1 {4\pi} (f'r -f)/\Sigma^2 , \label{rhof} \ee

and tangential and radial pressures

\be p_{rad}=-\rho, \quad p_{tan}=\rho - \frac 1 {8\pi}f''/\Sigma ,
\label{p}\ee

where $\Sigma=r^2.$ It yields for the interior the stress-energy
tensor $ T_{\mn} = \frac {3\alpha} {4\pi} diag (1,-1,-1,-1), $ or

\be\rho=-p_{rad}=-p_{tan}=\frac {3\alpha} {4\pi}, \label{rho} \ee

which generates a de Sitter interior for $\alpha >0$ and an anti de
Sitter interior for $\alpha <0$. If $\alpha =0, $ we have a flat
interior which corresponds to some previous classical models of an
electron, in particular, to the Dirac model of a charged sphere
and to the Lopez model in the form of a
 rotating elliptic shell \cite{Lop}.

The resulting sources may be considered as the bags filled by a
special matter with positive ($\alpha>0$) or negative
($\alpha < 0$)
energy density.\fn{It resembles the discussed at present structure
of dark energy and dark matter in Universe.
The case $\alpha >0$ is reminiscent
of the old Markov suggestions to consider particle as a semi-closed
Universe.}

The transfer from the external electro-vacuum solution to the
internal region (source) may be considered as a phase transition
from  `true' to `false' vacuum in a supersymmetric $U(1) \times \tilde
U(1) $ Higgs model \cite{BurBag}.

Assuming that transition region iii) is very thin, one can
consider the following graphical representation which turns out to
be very useful, see figure 2.

\begin{figure}[ht]
\centerline{\epsfig{figure=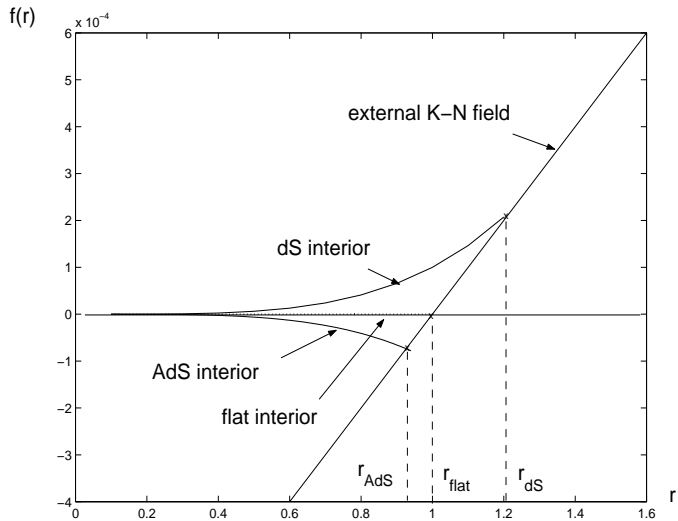,height=7cm,width=9cm}}
\caption{Regularization of the Kerr spinning particle by
matching the external field with  dS, flat or AdS interior.}
\end{figure}

The point of phase transition $r_0$ is determined by the equation
$f_{int}(r_0)=f_{KN}(r_0) ,$ which yields $\alpha r_0^4 = mr_0
-e^2/2 .$ From (\ref{rho}), we have $\rho=\frac {3\alpha} {4\pi}$
and obtain the equation

\be m= \frac {e^2} {2r_0} + \frac 4 3 \pi r_0^3 \rho . \ee

 In the first term on the right-hand side, one can easily recognize
 the electromagnetic mass of a charged sphere with radius $r_0$,
 $M_{em}(r_0)=\frac {e^2} {2r_0}$, while the second
 term is the mass of this sphere filled by a material with a
 homogenous
 density $\rho$, $M_m =\frac 4 3 \pi r_0^3 \rho .$
 Thus, the point of intersection $r_0$ acquires
a deep physical sense, providing an energy balance by the mass
formation. In particular, for the classical Dirac model
of a charged sphere with radius $r_0=r_e=\frac {e^2} {2m},$ the
balance equation yields the flat internal space with $\rho=0.$ If
$r_0=r_{dS} > r_e$, the interior is de Sitter space, and a material mass
of positive energy $M_m>0$ gives a
contribution to the total mass $m$. If $r_0 =r{AdS} < r_e ,$ this contribution
has to be negative $M_m <0 , $ which is accompanied by the
formation of an AdS internal space.

\section{Transfer to rotating case}

All the above treatments are valid for the rotating cases, and
for the passage to a rotating case, one has only to set

\be \Sigma=r^2 +a^2 \cos^2 \theta , \ee and consider $r$ and
$\theta$ as the oblate spheroidal coordinates \cite{BEHM}.
It looks wonderful, however it is a direct consequence of the structure
of function $H$, in which the nominator is independent from the rotation
parameter $a.$

The Kerr-Newman spinning particle with a spin $J=\frac 1 2 \hbar,$
 acquires the form of a relativistically rotating disk which
foliates on the rigidly rotating ellipsoidal shells, and
the board of the disk has $v\sim c$ \cite{BEHM}.
The corresponding
stress-energy tensor (\ref{rho}) describes in this case the matter
of source in a co-rotating with this disk coordinate system.
The disk has the form of a highly oblate
ellipsoid with thickness $r_0$ and radius $a=\frac 1 2 \hbar/m$
which is of order of the Compton length. Interior of the disk
represents a ``false'' vacuum having superconducting properties
\cite{Lop,BurBag},
so the charges are concentrated on the surface of this disk, at
$r=r_0$. Inside the disk, the local gravitational field is negligible.

\section{Nonstationarity and zero-point radiation.}

Classical models of a spinning particle encounter an unavoidable
conflicts with quantum theory.
The Kerr singular string acquires electromagnetic wave excitations
\cite{Bur,BurAxi,BurOri}. In classical theory these excitations lead
to a radiation which
breaks axial symmetry of the Kerr-Newman solution and leads to non-
stationarity. As a result, only an average metric takes the Kerr-Newman
form. In the Kerr-Schild formalism \cite{DKS},
electromagnetic excitations are related to a field
$\gamma(x)$ which induces electromagnetic radiation along the Kerr congruence $k_\m$
and non-stationarity of the solutions. This radiation leads also to infrared
divergence of the mass, and there are arguments that this radiation
has to be renormalized \cite{Bur,BurAxi,BurOri}, setting the field
$\gamma=0$.
In quantum theory oscillations are stationary and absence of
radiation caused by oscillations is postulated, although the radiation
is present in QED too, being related to radiative corrections: the field of
virtual photons, vacuum zero point field and vacuum polarization.

In a semiclassical approach, one can use the receipt of the quantum field theory in
curved spaces\cite{deWit}, which takes into account the quantum
effects concentrated in the divergent vacuum zero point field.
By the transfer to the classical Einstein-Maxwell theory, these quantum
vacuum fields have to be subtracted from the classical
stress-energy tensor  by a regularization  \cite{deWit}.

\be T^{(reg)}_{\mn} = T_{\mn} - <0|T_{\mn}|0>, \ee

which has to satisfy the condition

\be T^{(reg) \ \mn} ,_\m = 0 . \label{cons}\ee

It was conjectured in \cite{BurAxi,BurOri,BurNst} that regularization of the
Kerr-Newman stress-energy tensor has to be related with a subtraction of
electromagnetic radiation  caused by field $\gamma$ which
propagates along the Kerr congruence $k_\m$, and involves non-stationarity
by a loss of mass.
Twofoldedness of the Kerr geometry confirms this point of view, since
{\it the outgoing radiation on the `positive' out-sheet of the metric is
compensated by an ingoing radiation on the `negative' in-sheet}.
It shows, that the field $\gamma$ has to be
identified with the vacuum zero-point field and may be
subtracted from the stress-energy tensor by means of
regularization, which has to
satisfy the condition (\ref{cons}). Such regularization may
be performed, \cite{BurAxi,BurOri}, and leads to some modified
Kerr-Schild equations \cite{BurAxi}.
It shows that electromagnetic excitations on the Kerr background
are similar to the Casimir effect and may be
interpreted as a resonance of the zero-point fluctuations on the
(superconducting) source of the Kerr spinning particle
\cite{BurAxi,BurOri}.

 Although, the exact nontrivial solutions of the regularized system
have not been obtained so far, there were obtained corresponding
exact solutions of the Maxwell equations which show that any
`aligned' excitation of the Kerr geometry leads to the appearance of some
extra `axial' singular lines (strings)
which are semi-infinite and modulated by de Broglie
periodicity \cite{BurAxi,BurOri}.
 The recently obtained  multiparticle Kerr-Schild
solutions \cite{Multiks} support this point of view, leading to
the conclusion that the radiating twistorial structure of the
Kerr PNC belongs to the vacuum zero-point field, pointing out on
the twistorial texture of vacuum \cite{BurAxi,BurDir}.

{\bf Acknowledgements.} Author thanks the Organizing Committee
for kind invitation to this very interesting conference, and also
L. Pitaevskii and V. Ritus for very useful conversations.

{\bf Note added after publication:} a development of the presented
point of wiev is given in gr-qc/0606035, where we arrive at the conclusion
that the gravitational Kerr's description of
spinning particle may be dual to the QED
description, similar to the other dualities in superstring theory,
like AdS/CFT, strings/solitons and so on.

\end{document}